\documentclass[a4paper,12pt]{article}

\usepackage{url}
\usepackage{epsfig}
\usepackage{amsmath}
\usepackage{amssymb}
\usepackage{amsfonts}
\usepackage{bbm}
\usepackage[footnotesize]{caption}
\usepackage{graphicx}
\usepackage{mathrsfs}
\usepackage[font=scriptsize]{subcaption}
\usepackage{xcolor}
\usepackage{comment}
\usepackage{mathtools}
\usepackage{braket}
\usepackage{cite}
\usepackage{hyperref}
\usepackage{multirow}
\usepackage{relsize}
\usepackage{fullpage}
\usepackage{bbm}
\usepackage[T1]{fontenc}
\usepackage{pifont}
\usepackage{makecell}

\usepackage{graphicx} 
\usepackage{mathtools}

\newcommand{\pmatr}[1]{\begin{pmatrix} #1 \end{pmatrix}}

\newcommand{\newc}{\newcommand}
\newc{\be}{\begin{equation}}
\newc{\ee}{\end{equation}}
\newc{\bea}{\begin{eqnarray}}
\newc{\eea}{\end{eqnarray}}
\newc{\simlt}{~\mbox{\smaller\(\lesssim\)}~}
\newc{\simgt}{~\mbox{\smaller\(\gtrsim\)}~}

\begin{document}

\begin{titlepage}

\begin{center}
{\bf\Large  
Quark and Lepton Mass and Mixing with non-universal $Z'$ from 
a 5d Standard Model with gauged $SO(3)$
%a new $SO(3)$ isospin in 5d
} \\[12mm]
Francisco~J.~de~Anda$^{\dagger}$%
\footnote{E-mail: \texttt{fran@tepaits.mx}},
Stephen~F.~King$^{\star}$%
\footnote{E-mail: \texttt{king@soton.ac.uk}}
\\[-2mm]

\end{center}
\vspace*{0.50cm}
\centerline{$^{\star}$ \it
School of Physics and Astronomy, University of Southampton,}
\centerline{\it
SO17 1BJ Southampton, United Kingdom }
\vspace*{0.2cm}
\centerline{$^{\dagger}$ \it
Tepatitl{\'a}n's Institute for Theoretical Studies, C.P. 47600, Jalisco, M{\'e}xico}
\vspace*{1.20cm}

\begin{abstract}
{\noindent 
We propose a theory of quark and lepton mass and mixing with non-universal $Z'$ couplings based on 
a 5d Standard Model with quarks and leptons transforming as
triplets under a new gauged $SO(3)$ isospin. In the 4d effective theory, the $SO(3)$ isospin is broken to 
$U(1)'$, through a
$S^1/(\mathbb{Z}_2\times \mathbb{Z}'_2)$ orbifold, then subsequently dynamically
broken, resulting in a massive $Z'$. Quarks and leptons in the 5d bulk appear as massless modes, with zero Yukawa 
couplings to the Higgs on the brane, and zero couplings to $Z'$, at leading order, due to the $U(1)'$ symmetry. However, 
after the $U(1)'$ breaking, both 
Yukawa couplings and non-universal $Z'$ couplings are generated by heavy Kaluza-Klein exchanges.
Hierarchical quark and lepton masses result from a hierarchy of 5d Dirac fermion masses.
Neutrino mass and mixing arises from a novel type Ib seesaw mechanism, mediated by Kaluza-Klein Dirac neutrinos.
The non-universal  $Z'$ couplings may contribute to 
semi-leptonic $B$ decay ratios which violate $\mu - e$ universality. In this model such couplings 
are related to the corresponding quark and lepton effective Yukawa couplings. 
}
\end{abstract}

\end{titlepage}

\section{Introduction}
The flavour puzzle in the Standard Model (SM) is an indication that it is not complete.
Of the almost thirty parameters of the SM, most of them arise from unspecified Higgs Yukawa couplings. 
This provides a motivation for studying theories of flavour beyond the SM, in which the origin of Yukawa couplings 
is considered from a different perspective. 

One interesting approach that was proposed some time ago is based on the idea that 
the usual Higgs Yukawa couplings with the three families of chiral fermions are forbidden
by a discrete $Z_2$ symmetry, which is subsequently broken by some new scalar field $\langle \phi \rangle$, allowing 
the Higgs Yukawa couplings to arise effectively from mixing with a vector-like fourth 
family~\cite{Ferretti:2006df}. 
If the $Z_2$ symmetry is replaced by a 
gauged $U(1)'$ symmetry, under which the SM fermions are neutral, but the Higgs doublets are charged,
thereby forbidding Yukawa couplings but allowing mixing with the charged vector-like fourth family,
then after $U(1)'$ symmetry breaking, a massive $Z'$ gauge boson with non-universal couplings 
to quarks and leptons is generated by the mixing with the fourth family~\cite{King:2018fcg}.

In such a model~\cite{King:2018fcg}, the 
connection between non-universal $Z'$ couplings and the origin of Yukawa couplings may have interesting 
experimental implications. For example, some time ago, the LHCb Collaboration~\cite{Aaij:2014ora} 
and other experiments reported a number of anomalies in $B\rightarrow K^{(*)}l^+l^-$ 
decays such as the $R_K$ and $R_{K^*}$ ratios of $\mu^+ \mu^-$ to $e^+ e^-$ final states, 
which are observed to be about $80\%$ of their expected values with a $2.5 \sigma$ deviation from the SM.
Such anomalies may be accounted for by a new physics operator of the form~\cite{Hiller:2017bzc,Ciuchini:2017mik,Geng:2017svp,Capdevila:2017bsm,Ghosh:2017ber,Bardhan:2017xcc,DAmico:2017mtc,Descotes-Genon:2015uva,Calibbi:2015kma}  
$\bar b_L\gamma^{\mu} s_L \, \bar \mu_L \gamma_{\mu} \mu_L$,
with a coefficient $\Lambda^{-2}$ where $\Lambda \sim 30$ TeV.
This hints that there may be new physics arising from the non-universal couplings of 
leptoquarks and/or $Z'$ in order to generate such an operator.

This observation motivated many papers on non-universal $Z'$, many of them concerned with 
$U(1)'$ anomaly cancellation. In the above approach~\cite{King:2018fcg}, in which there are no chiral fermions which carry $U(1)'$
(only the $U(1)'$ charged vector-like fourth family) then anomaly cancellation is
automatic, and non-univerality is induced by mixing, making this 
a very natural and attractive possibility. Moreover, the resulting connection of the non-universal induced $Z'$ couplings
with Yukawa matrices~\cite{King:2018fcg}, is interesting.
However, such a model raises other questions, such as what is the origin of the vector-like fourth family with $U(1)'$ charges,
and why do they mix with the chiral quarks and leptons?

In this paper we take the next step in the development of such an approach to flavour, and suppose that 
the vector-like fourth family is identified as a Kaluza-Klein (KK) excitation of quarks and leptons which exist in a 5d bulk.
The origin of the $U(1)'$  gauge symmetry in such an approach is rather subtle since we require the chiral quarks and leptons to be $U(1)'$ isocharge zero  while their KK modes must include $U(1)'$ isocharged states. To address this, we shall suppose that the $U(1)'$ arises from a new gauged $SO(3)$ isospin in 5d, with the 5d quarks and leptons being assigned to $SO(3)$ isospin triplets, where each triplet decomposes into three states with $U(1)'$ isocharge $\pm 1, 0$. The idea is that the isocharge zero  state contains a massless mode, while the iso-charged states have only heavy KK modes. Such a framework has the nice feature that the scalar $\phi$ which breaks the $U(1)'$ and yields the $Z'$,
may originate as the fifth component of the 5d $SO(3)$ gauge field, according to a sort of ``gauge-Higgs'' unification \cite{Sakamura:2007qz,Medina:2007hz,Hall:2001zb,Gogoladze:2003bb,Hosotani:2006qp,Scrucca:2003ut,Antoniadis:2001cv,Panico:2005dh,Adachi:2018mby,Aranda:2019nac}.
In particular we shall follow the $SO(3)$ example in~\cite{Aranda:2019nac}, where the 
``gauge'' refers to the $SO(3)$ and the  ``Higgs'' refers to the $\phi$ scalar. 

In this way we are led to a simple ultraviolet completion of the vector-like family model
in which the quarks and leptons of the SM are extended into the 5d bulk,
while the SM Higgs remain as 4d brane fields. The hierarchies of Yukawa couplings are accounted for by assuming hierarchies
in the 5d quark and lepton Dirac masses, somewhat analogous to the way that Yukawa hierarchies are generated in a 5d Randall-Sundrum set-up with a warped extra dimension~\cite{Randall:1999ee,Huber:2000ie}, but here implemented with a flat extra dimension. However there are other important differences.
While small mass differences in Randall-Sundrum can lead to large hierarchies via their effect on the fermion wavefunction profiles,
here we require large 5d mass hierarchies which influence the Yukawa couplings directly, via seesaw type diagams, with the heavy fermions as messengers. The two different set-ups, with a warped or a flat extra dimension, are of course experimentally distinguishable, for example we do not predict a KK graviton here. Also the mass spectrum of the KK fermion modes is quite different.
And of course, here we will have a non-universal $Z'$ with experimental implications for flavour changing and non-universality.

The origin of neutrino masses in this set-up is also quite interesting since the usual 
type I seesaw mechanism cannot be implemented due to the absence of Majorana masses in 5d. Instead we shall rely on the recently proposed type Ib seesaw mechanism~\cite{Hernandez-Garcia:2019uof}, where heavy Dirac KK neutrino masses can yield small physical Majorana neutrino masses.

The outline of the remainder of the paper is as follows. In section~\ref{model} we define the 5d model, assuming zero 5d fermion masses, and show how Yukawa couplings originate. 
In section~\ref{masses} we show how the introduction of hierarchical 5d fermion masses can lead to Yukawa hierarchies and small CKM mixing angles. 
In section~\ref{neutrino} we show how neutrino Majorana masses can arise from KK Dirac neutrinos, via the type Ib seesaw mechanism.
In section~\ref{pheno} we discuss non-universal $Z'$ couplings and phenomenology.
Section~\ref{conclusion} concludes the main body of the paper. 
Appendix~\ref{app:orb} details the $S^1/(\mathbb{Z}_2\times \mathbb{Z}'_2)$ orbifold.
Appendix~\ref{app:so} discusses the $SO(3)$ decomposition.
Appendix~\ref{app:h} defines the Higgs fields on the branes and shows how the 
$U(1)'$ neutral Higgs components gain large masses.

\section{The 5d Standard Model with gauged $SO(3)$ isospin}
\label{model}

Suppose that the Standard Model (SM) is extended into a flat 5d spacetime, where each 5d quark and lepton field is
assumed to be an isotriplet under a new gauged $SO(3)$ isospin.
This is not weak isospin, nor is it a family symmetry, it is a completely new degree of freedom carried by each 5d multiplet
$Q^{\alpha}_i,u^{\alpha}_i,d^{\alpha}_i,L^{\alpha}_i,e^{\alpha}_i$, where $i=1,2,3$ is a family index and $\alpha=1,2,3$ is a new
$SO(3)$ index.
The extra dimension $y$ is orbifolded as $S^1/(\mathbb{Z}_2\times \mathbb{Z}'_2)$, resulting in two 4d branes, one at $y=0$
and the other at $y=\pi R/2$, connected via the 5d bulk (see Appendix~\ref{app:orb} for details). The quarks and leptons live in the bulk in irreducible representations of the 5d Lorentz group $SO(1,4)$ which is broken to the standard one $SO(1,3)$. The fields decompose into the standard representations, as described in Appendix~\ref{app:orb}.
We also assume that the SM gauge symmetry $SU(3)_C\times SU(2)_L\times U(1)_Y$
lives in the bulk, together with the gauged isospin symmetry $SO(3)$, while the Higgs doublets live on the 4d branes. We assume the existence of two physical Higgs doublets as in \cite{King:2018fcg}. The $\mathbb{Z}_3$ symmetry enforces the $H_{u,d}$ couplings corresponding to up-type and down-type fermions respectively, but allows the Neutrino Majorana mass terms.

This extended symmetry, 5d Lorentz and $SO(3)$ gauged isospin, is broken 
with independent $SO(3)$ boundary conditions at each brane as follows:
\begin{equation}
P_0=\mathbb{I},\ \ \ P_{\pi R/2}=diag (-1,1,-1),
\end{equation}
where the $SO(3)$ decomposition under $U(1)'$ is detailed in Appendix~\ref{app:so}, the 
key result being that each $SO(3)$ isotriplet decomposes into three states with $U(1)'$ isocharges $+1,0,-1$.
The $SO(3)$ gauge fields in the bulk have the boundary conditions shown in Table~\ref{ta:fcq}, which leads to a preserved $U(1)'$ 4d massless gauge boson, plus extra KK gauge boson excitations, plus the fifth scalar components of the gauge fields which are 
all heavy KK modes. We also assume a complex scalar in the  brane which acquires a VEV and breaks the gauged $U(1)'$.
The SM fermions, which live in the bulk as $SO(3)$ triplets, are assigned a parity under each boundary condition as shown
in Table~\ref{ta:fcq}, leading to massless modes consisting of the usual three chiral families, which are all neutral under the $U(1)'$
(i.e. have isocharge zero)
plus their KK excitations (both charged and neutral under $U(1)'$).
We also introduce a 5d neutrino field which transforms as a SM and $SO(3)$ singlet, with the boundary conditions shown in Table~\ref{ta:fcq}, such that no massless modes are present; these heavy KK neutrinos play the role of ``right-handed neutrinos'' in the seesaw mechanism, although in this case, being KK modes, they are heavy Dirac fermions (see later).

\begin{table}
\centering
\begin{tabular}{c|ccccc|c  c c }
 \hline
   \hline
Field &$SU(3)_C$ & $SU(2)_L$ & $U(1)_Y$ & $SO(3)$ & $\mathbb{Z}_3$  & 4d field &  $P_0\mathcal{P}_5$  & $P_{\pi R}\mathcal{P}_5$\\
\hline
$\textbf{Q}_i$ &   $\textbf{3}$ & $\textbf{2}$ & $1/6$ &$\textbf{3}$ &$1$ & $Q_{iL}^\pm$  &$1$ & $-1$   \\
  &  &&&& &$Q_{iL}^0$ &$1$ &  $1$  \\
 &  &&&&& $Q_{iR}^\pm$  &$-1$ & $1$  \\
 &  &&&& &$Q_{iR}^0$ &$-1$ & $-1$  \\
  $\textbf{u}_i$ &$\textbf{3}$ & $\textbf{1}$ & $2/3$& $\textbf{3}$  &$\omega$& $u_{iR}^{\pm}$  &$1$ & $-1$   \\
  &  &&&&& $u_{iR}^{0}$ &$1$ &  $1$  \\
 &  &&&&& $u_{iL}^{\pm}$  &$-1$ & $1$  \\
 &  &&&&& $u_{iL}^{0}$ &$-1$ & $-1$  \\
 $\textbf{d}_i$ & $\textbf{3}$ & $\textbf{1}$ & $-1/3$ & $\textbf{3}$  &$\omega$& $d_{iR}^{\pm}$  &$1$ & $-1$   \\
  &  &&&&& $d_{iR}^{0}$ &$1$ &  $1$  \\
 &  &&&&& $d_{iL}^{\pm}$  &$-1$ & $1$  \\
 &  &&&&& $d_{iL}^{0}$ &$-1$ & $-1$  \\
 $\textbf{L}_i$ &   $\textbf{1}$ & $\textbf{2}$ & $-1/2$ &$\textbf{3}$ &$1$ & $L_{iL}^\pm$  &$1$ & $-1$   \\
  &  &&&& &$L_{iL}^0$ &$1$ &  $1$  \\
 &  &&&&& $L_{iR}^\pm$  &$-1$ & $1$  \\
 &  &&&& &$L_{iR}^0$ &$-1$ & $-1$  \\
 $\textbf{e}_i$ & $\textbf{1}$ & $\textbf{1}$ & $-1$ & $\textbf{3}$  &$\omega$& $e_{iR}^{\pm}$  &$1$ & $-1$   \\
  &  &&&&& $e_{iR}^{0}$ &$1$ &  $1$  \\
 &  &&&&& $e_{iL}^{\pm}$  &$-1$ & $1$  \\
 &  &&&&& $e_{iL}^{0}$ &$-1$ & $-1$  \\
 \hline
  $\boldsymbol\nu$ &$\textbf{1}$ & $\textbf{1}$ & $0$& $\textbf{1}$  &$\omega$& $\nu_R^{c0}$  &$1$ & $-1$   \\
 &  &&&&& $\nu_L^{c0}$ &$-1$ & $1$  \\
\hline
$\textbf{A}_M$  & $\textbf{1}$ &$\textbf{1}$ & 0&$\textbf{3}$  &$1$& $A_\mu^0$  &$1$ & $ 1$  \\
 &&&&&& $A_5^0$  &$-1$ & $-1$   \\
  &&&&&& $A_\mu^\pm$  &$1$ & $ -1$  \\
   &&&&&& $A_5^\pm$  &$-1$ & $1$  \\
    \hline \hline
  $\boldsymbol H_{u,d}$ &$\textbf{1}$ & $\textbf{2}$ & $\mp 1/2$& $\textbf{3}$  &$\omega^2$& $H_{u,d}^{0,\pm}$  &$-$ & $-$   \\
  $\boldsymbol H'_{u,d}$ &$\textbf{1}$ & $\textbf{2}$ & $\mp 1/2$& $\textbf{1}$  &$\omega^2$& $H_{u,d}^{0'}$  &$-$ & $-$   \\
   \hline
   \hline
  \end{tabular}
\caption{\label{ta:fcq}The field content of the model, comprising three 5d SM fermion families $\textbf{Q}_{i}$, $\textbf{u}_{i}$,
$\textbf{d}_{i}$, $\textbf{L}_{i}$,  $\textbf{e}_{i}$, plus one singlet neutrino $\boldsymbol\nu$, 
plus the 5d $SO(3)$ gauge field $\textbf{A}_M$,
with their decomposition under the orbifold breaking. The index $i=1,2,3$ is a family index not an 
$SO(3)$ index, i.e. there are three families of $SO(3)$ isotriplets. 
When the $SO(3)$ is broken to a $U(1)'$ by the orbifold, 
this will yield a massless neutral fermion 
(under the $U(1)'$) for each fermion family. 
There is also an $U(1)'$ charged scalar $\phi$ in the $\pi R/2$ brane, whose VEV
will eventually break the $U(1)'$, yielding a massive $Z'$. As it is an incomplete $SO(3)$ multiplet, it is not shown in this table.
The 5d SM gauge fields are not displayed explicitly here.
 The 4d scalar
Higgs doublets on the branes are also shown, with $\boldsymbol H_{u,d}$ on the zero brane and $\boldsymbol H'_{u,d}$ on the other brane.}
\end{table}

Each bulk field can be expanded as a sum of their modes \cite{Fujimoto:2016llj,Kawamura:2000ev}. 
First let us decompose the $SO(3)$ gauge vector $A_M$ field,
\begin{equation}\begin{split}
\textbf{A}_M(x,y)&\to  \textbf{A}_\mu^0(x,y)= \sqrt{2 \over {\pi R}} 
      \sum_{n=0}^{\infty} (A_\mu^0)^{(2n)}(x) \cos{2ny \over R} 
      \\ &\quad \ \textbf{A}_\mu^\pm(x,y)=\sqrt{2 \over {\pi R}} 
      \sum_{n=0}^{\infty} (A^\pm_\mu)^{(2n+1)}(x) \cos{(2n+1)y \over R} \\
      &\quad \ \textbf{A}_5^0(x,y)=\sqrt{2 \over {\pi R}}
      \sum_{n=0}^{\infty} (A^0_5)^{(2n+2)}(x) \sin{(2n+2)y \over  R} \\
      &\quad \ \textbf{A}_5^\pm(x,y)= \sqrt{2 \over {\pi R}}
      \sum_{n=0}^{\infty} (A_5^\pm)^{(2n+1)}(x) \sin{(2n+1)y \over  R} .
\end{split}\end{equation}

The quark and lepton electroweak doublets ($F=Q,L$) (dropping the flavour index $i$ and the $SO(3)$ index $\alpha$) are decomposed in 4d as,
\begin{equation}
\begin{split}
\textbf{F}(x,y)&\to \textbf{F}_L^\pm(x,y)=\sqrt{2 \over {\pi R}} 
      \sum_{n=0}^{\infty} (F_L^\pm)^{(2n+1)}(x) \cos{(2n+1)y \over R}, \\
      &\quad \ \textbf{F}_L^0(x,y)=\sqrt{2 \over {\pi R}} 
      \sum_{n=0}^{\infty} (F_L^0)^{(2n)}(x) \cos{2ny \over R},\\
      &\quad \ \textbf{F}_R^\pm(x,y)=\sqrt{2 \over {\pi R}}
      \sum_{n=0}^{\infty} (F_R^\pm)^{(2n+1)}(x) \sin{(2n+1)y \over  R},\\
      &\quad \ \textbf{F}_R^0(x,y)= \sqrt{2 \over {\pi R}}
      \sum_{n=0}^{\infty} (F_R^0)^{(2n+2)}(x) \sin{(2n+2)y \over  R},
      \label{eq:5dfer}
\end{split}
\end{equation}
where the isoneutral left-handed doublets  ($F_L^0=Q_L^0,L_L^0$) clearly have zero modes for $n=0$.
The quark and lepton electroweak singlets ($f=u,d,e$) (again dropping the flavour index $i$ and the $SO(3)$ index $\alpha$) 
are chosen to have the opposite parity at the zero brane, as compared to the doublets above,
so their 4d profiles are,
\begin{equation}
\begin{split}
\textbf{f}(x,y)&\to \textbf{f}_R^\pm(x,y)=\sqrt{2 \over {\pi R}} 
      \sum_{n=0}^{\infty} (f_R^\pm)^{(2n+1)}(x) \cos{(2n+1)y \over R}, \\
      &\quad \ \textbf{f}_R^0(x,y)=\sqrt{2 \over {\pi R}} 
      \sum_{n=0}^{\infty} (f_R^0)^{(2n)}(x) \cos{2ny \over R},\\
      &\quad \ \textbf{f}_L^\pm(x,y)=\sqrt{2 \over {\pi R}}
      \sum_{n=0}^{\infty} (f_L^\pm)^{(2n+1)}(x) \sin{(2n+1)y \over  R},\\
      &\quad \ \textbf{f}_L^0(x,y)= \sqrt{2 \over {\pi R}}
      \sum_{n=0}^{\infty} (f_L^0)^{(2n+2)}(x) \sin{(2n+2)y \over  R},
      \label{eq:5dfel}
\end{split}
\end{equation}
so that isoneutral right-handed singlets ($f^0_R=u^0_R,d^0_R,e^0_R$) have massless zero modes for $n=0$.
One neutrino field, a singlet under all gauge groups, decomposes in 4d as,
\begin{equation}
\begin{split}
\boldsymbol\nu (x,y)&\to \boldsymbol\nu_L (x,y)=\sqrt{2 \over {\pi R}} 
      \sum_{n=0}^{\infty} (\nu_L)^{(2n+1)}(x) \cos{(2n+1)y \over  R} \\
      &\quad\  \boldsymbol\nu_R(x,y)=\sqrt{2 \over {\pi R}}
      \sum_{n=0}^{\infty} (\nu_R)^{(2n+1)}(x) \sin{(2n+1)y \over  R},
\end{split}
\end{equation}
so that the zero modes form a single massive Dirac KK state $(\nu_L,\nu_R)$, as indeed do all the higher KK modes.
Indeed all the heavy KK modes of all the fermions pair up into massive Dirac states in a similar way, but it is most simply seen with the neutrinos which have no massless modes (unlike the other electrically charged fermions which all contain massless modes as well as heavy KK modes).

The Higgs doublets are located on the 4d branes, so they will not have any KK mode expansions.
We desire two Higgs doublets $H_{u,d}$ to be located at the zero brane, each with a $U(1)'$ isocharge of $-1$,
which would forbid usual Higgs Yukawa couplings (since the massless chiral quarks and leptons are neutral),
but which will allow effective Yukawa couplings to be generated when the $U(1)'$ is broken, via their coupling to 
isocharged KK states, as discussed later. In order to achieve this, it is actually necessary to introduce complete
$SO(3)$ triplets of Higgs $H_{u,d}$ at the zero brane (where the $SO(3)$ is unbroken) plus an additional a pair of Higgs doublets
$H'^0_{u,d}$ at the other brane, which are neutral under $U(1)'$, and serve to give a large mass to the unwanted neutral components of the triplet at the zero brane. This mechanism is discussed in detail in Appendix~\ref{app:h}.

\begin{figure}
 \centering
\includegraphics[scale=0.51]{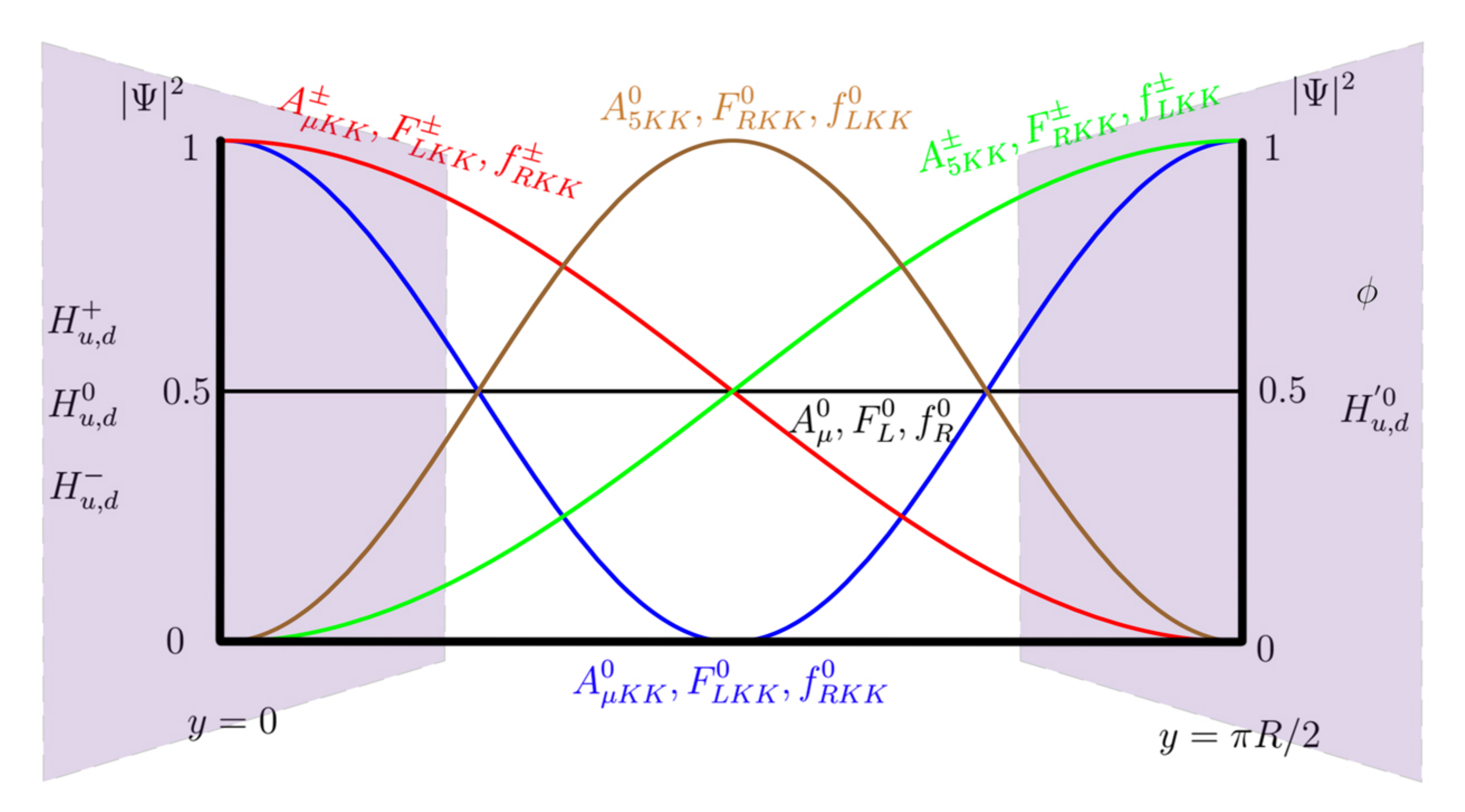}
		\caption{Fixed branes (locating the Higgs) at $y=0, \pi R/2$ and the normalised wavefunction squared $|\psi |^2$ profiles for the lowest bulk modes. The $H^-_{u,d}$ are the physically relevant Higgs doublets renamed $H_{u,d}$ in the text.
		The massless $n=0$ modes $A^0_{\mu}, F^0_L,f^0_R$ are isoneutral (isocharge zero) and correspond to the horizontal black line at $|\psi |^2=0.5$. Their isoneutral $n=1$ KK modes have square wavefunctions depicted by the blue curve. Other $n=0$ isoneutral modes are massive KK states shown by the brown curve. The $n=0$ isocharged square wavefunctions are massive KK modes indicated by the green, red curves which vanish on the branes at $y=0, \pi R/2$, respectively.}
		\label{fig:prof}
\end{figure}

The 5d kinetic terms for these fields, including the Higgs $H_{u,d}$ couplings at the zero brane, which together generate all the fermion masses~\footnote{The mass dimensionality of each of the 5d fields are $[A_M]=3/2$ for gauge vectors\\
, $[F]=2$ for fermions and $[H]=1$ for brane scalars.},
\begin{equation}
\begin{split}
\mathcal{L}_{m\psi}=&
i\bar{\textbf{Q}}_i\gamma^M D_M\textbf{Q}_i+i\bar{\textbf{u}}_i\gamma^MD_M \textbf{u}_i+i\bar{\textbf{d}}_i\gamma^M D_M\textbf{d}_i\\
&+i\bar{\textbf{L}}_i\gamma^M D_M\textbf{L}_i+i\bar{\textbf{e}}_i\gamma^MD_M \textbf{e}_i+i\bar{\boldsymbol{\nu}}\gamma^M D_M\boldsymbol{\nu}
\\
& +\delta(y)\Big[ D_\mu H_u (D^\mu H_u)^\dagger+D_\mu H_d (D^\mu H_d)^\dagger-V(H_u,H_d)\Big]
\\
& +\delta(y-\pi R/2)\Big[ (D^\mu\phi)^\dagger D_\mu\phi-V(\phi)\Big]
\\
&+\frac{\delta(y)}{\Lambda}\Big[y_u^{ij}\textbf{Q}_i^\dagger H_u \textbf{u}_j+y_d^{ij}\textbf{Q}_i^\dagger H_d \textbf{d}_j+y_e^{ij}\textbf{L}_i^\dagger\textbf{H}_d \textbf{e}_j+y_\nu^i\textbf{L}_i^\dagger H_u \boldsymbol\nu+y^{'j}_\nu \textbf{L}_j^\dagger\tilde{H}_d \boldsymbol\nu^\dagger\Big]
\\
&+\frac{\delta(y-\pi R/2)}{\Lambda}\phi(Y_Q^{ij} \bar{\textbf{Q}}_i\textbf{Q}_j+Y_u^{ij} \bar{\textbf{u}}_i\textbf{u}_j+Y_d^{ij} \bar{\textbf{d}}_i\textbf{d}_j+Y_L^{ij} \bar{\textbf{L}}_i\textbf{L}_j+Y_e^{ij} \bar{\textbf{e}}_i\textbf{e}_j+Y_\nu^{ij} \bar{\boldsymbol\nu}_i\boldsymbol\nu_j)
\label{eq:5dy}
\end{split}
\end{equation}
where the $y^{ij}$ are flavour matrices. The Higgses and the $\phi$ have their potential, which we do not explicitly show. The potential drives the VEV of the fields. We assume that the $\phi$ gets a larger VEV, as it is in the $SO(3)$ broken brane, we can assume it only has a positively charged $U(1)'$ component. The Higgs potential will have the the usual hierarchy and stability problems, which are not addressed in this model.

We now turn to the generation of the Yukawa couplings.
This is non-trivial since the massless
mode quarks and leptons have zero $U(1)'$ charge, and are hence forbidden to couple to the 4d Higgs $H_{u,d}$,
which are charged under $U(1)'$, since the neutral Higgs components gain large masses by the mechanism in 
Appendix~\ref{app:h}, and so will not develop VEVs.
The 4d effective theory relevant for the generation of Yukawa couplings 
will consist of the massless quark and lepton modes coupling to the 4d Higgs fields and the heavy KK modes.
The SM field content lies in the zero modes, where we relabel the physically relevant fields as in Table~\ref{ta:ren}. 
Then the terms relevant for the generation of effective Yukawa couplings involve the fields in Table~\ref{ta:ren}.

\begin{table}
\centering
\begin{tabular}{c|ccccc|c   }
 4d field &$SU(3)_C$ & $SU(2)_L$ & $U(1)_Y$ & $U(1)'$ & $\mathbb{Z}_3$  &  mode origin\\
\hline
$Q_{iL}$ &   $\textbf{3}$ & $\textbf{2}$ & $1/6$ &$0$ &$1$ & $(Q_{iL}^0)^0$     \\
  $u_{iR}$ &$\textbf{3}$ & $\textbf{1}$ & $2/3$& $0$  &$\omega$& $(u_{iR}^{0})^0$    \\
 $d_{iR}$ & $\textbf{3}$ & $\textbf{1}$ & $-1/3$ & $0$  &$\omega$& $(d_{iR}^{0})^0$    \\
 $e_{iR}$ & $\textbf{1}$ & $\textbf{1}$ & $-1$ & $0$  & $\omega $ & $(e_{iR}^{0})^0$  \\
  $L_{iL}$ &   $\textbf{1}$ & $\textbf{2}$ & $-1/2$ &$0$ &$1$ & $(L_{iL}^0)^0$    \\
\hline
$Q_{iLKK}$ &   $\textbf{3}$ & $\textbf{2}$ & $1/6$ &$-1$ &$1$ & $(Q_{iL}^-)^{(2n+1)}$     \\
$Q_{iRKK}$ &   $\textbf{3}$ & $\textbf{2}$ & $1/6$ &$-1$ &$1$ & $(Q_{iR}^-)^{(2n+1)}$     \\
  $u_{iRKK}$ &$\textbf{3}$ & $\textbf{1}$ & $2/3$& $-1$  &$\omega$& $(u_{iR}^{-})^{(2n+1)}$    \\
   $u_{iLKK}$ &$\textbf{3}$ & $\textbf{1}$ & $2/3$& $-1$  &$\omega$& $(u_{iL}^{-})^{(2n+1)}$    \\
 $d_{iRKK}$ & $\textbf{3}$ & $\textbf{1}$ & $-1/3$ & $-1$  &$\omega$& $(d_{iR}^{-})^{(2n+1)}$    \\
 $d_{iLKK}$ & $\textbf{3}$ & $\textbf{1}$ & $-1/3$ & $-1$  &$\omega$& $(d_{iL}^{-})^{(2n+1)}$    \\
  $e_{iRKK}$ & $\textbf{1}$ & $\textbf{1}$ & $-1$ & $-1$  &$\omega$& $(e_{iR}^{-})^{(2n+1)}$  \\
  $e_{iLKK}$ & $\textbf{1}$ & $\textbf{1}$ & $-1$ & $-1$  &$\omega$& $(e_{iL}^{-})^{(2n+1)}$  \\
  $L_{iLKK}$ &   $\textbf{1}$ & $\textbf{2}$ & $-1/2$ &$-1$ &$1$ & $(L_{iL}^-)^{(2n+1)}$    \\
  $L_{iRKK}$ &   $\textbf{1}$ & $\textbf{2}$ & $-1/2$ &$-1$ &$1$ & $(L_{iR}^-)^{(2n+1)}$    \\
  \hline
  $L'_{iLKK}$ &   $\textbf{1}$ & $\textbf{2}$ & $-1/2$ &$+1$ &$1$ & $(L_{iL}^+)^{(2n+1)}$    \\
  $L'_{iRKK}$ &   $\textbf{1}$ & $\textbf{2}$ & $-1/2$ &$+1$ &$1$ & $(L_{iR}^+)^{(2n+1)}$    \\
  \hline
  $\nu_{LKK}$ & $\textbf{1}$ & $\textbf{1}$ & $0$ & $0$  &$\omega$& $(\nu_L)^{(2n+1)}$  \\
    $\nu_{RKK}$ & $\textbf{1}$ & $\textbf{1}$ & $0$ & $0$  &$\omega$& $(\nu_R)^{(2n+1)}$  \\
 \hline
$Z'_\mu$  & $\textbf{1}$ &$\textbf{1}$ & 0&$0$  &$1$& $(A_\mu^0)^0$    \\
\hline
$H_u$ & $\textbf{1}$ &$\textbf{2}$ & $-1/2$&$-1$  &$\omega^2$& 4d brane   \\
 $H_d$ & $\textbf{1}$ &$\textbf{2}$ & $1/2$&$-1$  &$\omega^2$& 4d brane   \\
$\phi$  & $\textbf{1}$ &$\textbf{1}$ & 0&$+ 1$  &$1$& 4d brane
\end{tabular}
\caption{\label{ta:ren} Field content in the effective 4d theory and their mode origin in the 5d expansion. 
All 4d fermion fields are left-handed Weyl spinors but they originate from 5d Dirac fermion modes with left (L) and right (R) handed components
as shown in the last column (where the R components have been CP conjugated to yield left-handed Weyl spinors).
The left-handed SM fermion states without a KK subscript correspond to the lowest Kaluza Klein modes having zero 
KK mass contributions in the 4d effective theory.
We only show the KK fermion modes with positive $U(1)'$ charge since they play a role in generating quark and lepton Yukawa couplings.
There will be other KK fermion modes (not displayed) of all three isocharges $\pm 1, 0$, arising from $SO(3)$ triplets, which  
similarly form massive vector-like pairs of left-handed Weyl spinors with conjugate quantum numbers.
For example, we display $L'_{iKK}$, which has negative $U(1)'$ isocharge, since it plays a role in the type Ib seesaw mechanism.
The neutrinos $\nu_{KK}$ arising from $SO(3)$ singlets do not have any massless modes.
The spin-1 $Z'_\mu$ field  originate from the 5d $SO(3)$ gauge field. The $\phi$ scalar is located in the $\pi R/2$ brane.
The two scalar Higgs doublets $H_u\equiv H_u^-$, $H_d\equiv H_d^-$ are the isocharge negative 4d brane fields
arising from the isotriplet Higgs on the zero brane.}
\end{table}

The couplings relevant for the generation of effective zero mode quark Yukawa couplings involving the fields in Table~\ref{ta:ren},
are given by,
\begin{equation}
\begin{split}
&\mathcal{L}_{0Q}= 
 \left(\frac{2}{\pi R}\right)\frac{1}{\Lambda}\sum_n\int_0^{\pi R/2} dy\ \delta(y)\cos{(2n+1)y \over R} \left[ y_u^{ij} (Q_{iLKK}^\dagger H_u u_{jR}+Q_{iL}^\dagger H_u u_{jRKK})\right]\\
&+\left(\frac{2}{\pi R}\right)\frac{1}{\Lambda} \int_0^{\pi R/2}dy \ \delta(y-\pi R/2)\   \sin \frac{(2n+1)y}{R}    \left[ Y_Q^{ij}  Q^{\dagger}_{iL} \phi Q_{jRKK}+ Y_u^{ij} u^{\dagger}_{iLKK}\phi u_{jR}\right]\\ \\
&\ \ \ \  + u\leftrightarrow d.
\end{split}
\end{equation}
For simplicity, we can just work with the $n=0$ mode and integrate out the 5th dimension,
\begin{equation}
\begin{split}
\mathcal{L}_{0Q}&= 
\frac{2}{\pi R\Lambda} \Big[ y_u^{ij} (Q_{iLKK}^\dagger H_u u_{jR}+Q_{iL}^\dagger H_u u_{jRKK}) \Big]
\\
&+ \left(\frac{2}{\pi R\Lambda }\right) \Big[ Y_Q^{ij}  Q^{\dagger}_{iL} \phi Q_{jRKK}+Y_u^{ij}  u^{ \dagger}_{iLKK}\phi u_{jR}\Big]\\
&+u\leftrightarrow d.
\label{L0Q}
\end{split}
\end{equation}
The charged lepton couplings are built in a similar way, leading to 
\begin{equation}
\begin{split}
\mathcal{L}_{0L}&= 
\frac{2}{\pi R\Lambda} \Big[ y_e^{ij} (L_{iLKK}^\dagger H_d e_{jR}+L_{iL}^\dagger H_d e_{iRKK}) \Big]\\
&+ \left(\frac{2}{\pi R\Lambda}\right) \Big[  Y_L^{ij} L^{\dagger}_{iL} \phi L_{jRKK}+ Y_e^{ij}  e^{ \dagger}_{iLKK}\phi e_{jR}\Big].
\end{split}
\label{L0L}
\end{equation}

\begin{figure}[h]
\centering
\includegraphics[scale=0.25]{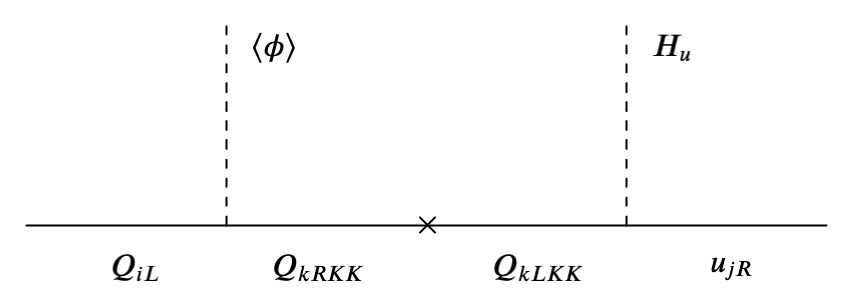}
\includegraphics[scale=0.25]{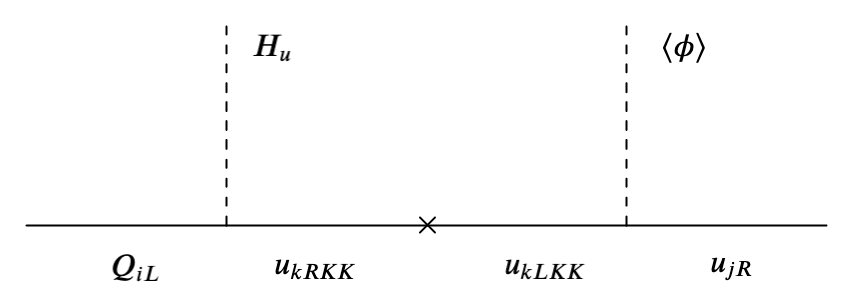}
		\caption{Diagrams for effective 4d up-type quark Yukawa couplings (see Table~\ref{ta:ren} for notation). Similar diagrams with $u\rightarrow d$ give the down-type Yukawa couplings. Charged lepton Yukawa couplings arise from similar diagrams involving $L_i$, $e_j$, together with their respective 
		leptonic KK messengers,
		and $H_d$.}
		\label{fig:YF}
\end{figure}

By integrating out the mediators we obtain the effective Yukawa terms
\begin{equation}
\begin{split}
\mathcal{L}_Y= \frac{8}{\pi \Lambda}\left(\frac{2}{R\Lambda\pi}\right)\braket{\phi}&\Big[(Y_Q^{ik}y_u^{kj}+y_u^{ik}Y_u^{kj})   Q_{iL}^\dagger H_u u_{jR}+(Y_Q^{ik}y_d^{kj}+y_d^{ik}Y_d^{kj}) Q_{iL}^\dagger H_d d_{jR}\\
&+(Y_L^{ik}y_e^{kj}+y_e^{ik}Y_e^{kj}) L_{iL}^\dagger H_d e_{jR}\Big].
\label{eq:effyuk}
\end{split}
\end{equation}
The quarks and charged lepton Yukawa couplings are mediated by each right and left handed fermion, as well as the Higgs KK modes, as shown in figure \ref{fig:YF}.
Unfortunately, the theory so far does not provide any understanding of the quark and lepton mass hierarchies, it simply reparametrises the Standard Model Yukawa couplings in terms of the 5d Yukawa couplings. This deficiency is remedied in the next section.

The strongest constraint on the compactification scale comes from LEP constraints on Flavour Changing Neutral Currents generated by KK gluons as well as Precision Electroweak Tests modified by KK $W$ and $Z$ \cite{Cheung:2001mq,Barbieri:2004qk}
\begin{equation}
\frac{1}{R}>6\ {\rm TeV},
\label{eq:comp}
\end{equation}
which has to be complied in any further calculation.

\section{Explicit 5d masses and Yukawa hierarchies}
\label{masses}
So far we have been ignoring the effect of the 5d scalar and fermion masses, and the analysis in the previous section 
implicitly assumes these masses to be zero.
However the 5d fermions are always Dirac fermions, therefore it is possible to write explicit mass terms for each 5d fermion. 
Similarly we can include 5d scalar Higgs doublet masses. Therefore  in addition to the terms in eq. \ref{eq:5dy}, one can write explicit 5d 
masses as
\begin{equation}
\begin{split}
\mathcal{L}_{5dm}=&\bar{\textbf{Q}}_i M^{ij}_Q \textbf{Q}_j+\bar{\textbf{u}}_i M^{ij}_u \textbf{u}_j+\bar{\textbf{d}}_iM^{ij}_d\textbf{d}_j
+\bar{\textbf{L}}_i M^{ij}_L\textbf{L}_j+\bar{\textbf{e}}_iM^{ij}_e \textbf{e}_j+\bar{\boldsymbol{\nu}}M_{\nu}\boldsymbol{\nu},
\label{eq:5m}
\end{split}
\end{equation}
where we can diagonalize the matrix from the start so that $M^{ij}=M^i\delta^{ij}$.

We assume that these mass parameters are above the compactification scale. As these are Dirac masses, the chiral zero modes are unaffected. However, if there where scalars in the bulk, this mass would affect the zero mode. To avoid fine tuning on these parameters, all the scalars have been assumed on the brane.

These 5d masses will change the spectrum of the KK masses. It does not change the mode profiles in eq. \ref{eq:5dfer}. However the addition of the new mass parameters changes the mass spectrum of the KK modes. 
However, the zero modes remain massless~\cite{Ponton:2012bi}. The scalar and fermion massive KK modes change their mass as \cite{Scrucca:2003ra}
 \begin{equation}
 \frac{1}{R}\to \sqrt{R^{-2}+M_i^2}.
 \end{equation}
 The effective Yukawa terms become 
 \begin{equation}
\begin{split}
\mathcal{L}_Y=&\frac{4g}{\pi }\left(\frac{2}{\pi R\Lambda}\right) \braket{\phi}\times \\
&\left[\left(\pi R/2+\big(R^{-2}+(M_{Q}^i)^2\big)^{-1/2} Y^{ik}_Q y_u^{kj} +\big(R^{-2}+(M_u^j)^2\big)^{-1/2}y_u^{ik} Y^{kj}_u\right)  Q_{iL}^\dagger H_u u_{jR}\right.\\
&\quad \quad+\left(\pi R/2+\big(R^{-2}+(M_{Q}^i)^2\big)^{-1/2}Y^{ik}_Q y_d^{kj}+\big(R^{-2}+(M_d^j)^2\big)^{-1/2}y_d^{ik}Y^{kj}_d\right)  Q_{iL}^\dagger H_d d_{jR}\\
&\quad \quad+\left.\left(\pi R/2+\big(R^{-2}+(M_{L}^i)^2\big)^{-1/2}Y^{ik}_L y_e^{kj} +\big(R^{-2}+(M_e^j)^2\big)^{-1/2}y_e^{ik}Y^{kj}_e\right) L_{iL}^\dagger H_d e_{jR}\right].
\label{Yuks}
\end{split}
\end{equation}

Note that the relative contributions of each fermion and Higgs mediated diagram has changed, depending on the values of the 5d masses.
Let us make the following assumptions about the 5d masses:
\begin{itemize}
\item one of the three fermion families (say the third family) has the lightest 5d mass, 
as compared to the first two families, $M_3\ll M_{1,2}$
\item within the third fermion family, the hierarchies of 5d masses satisfy,
$1/R \lesssim M_{Q_3}\ll M_{d_3} \ll M_{u_3}$ and $1/R \lesssim M_{L_3}\ll M_{e_3}$
\end{itemize}

With these assumptions the dominant contributions to the Yukawa couplings arise from the left diagram in 
figure \ref{fig:YF} which is mediated by the third family doublets $Q_{3KK}$ in the case of quarks, or
$L_{3KK}$ in the case of the charged leptons. 
The interactions involving the lightest $Q_{3KK}$ and $L_{3KK}$ states are just a subset of the terms in Eqs.\ref{L0Q},
\ref{L0L},
\begin{equation}
\begin{split}
\mathcal{L}_{Q_{3KK}}&= 
 \frac{2}{\pi R\Lambda} \Big[ y_u^{3j} (Q_{3LKK}^\dagger H_u u_{jR}) +u\leftrightarrow d\Big]
+\left(\frac{2}{R\Lambda\pi }\right)Y^{33}_Q \Big[  Q^{\dagger}_{3L} \phi Q_{3RKK}\Big],
\end{split}
\end{equation}
\begin{equation}
\begin{split}
\mathcal{L}_{L_{3KK}}&= 
\frac{2}{\pi R\Lambda}  \Big[ y_e^{3j} (L_{3LKK}^\dagger H_d e_{jR}) \Big]
+\left(\frac{2}{R\Lambda\pi }\right) \Big[  L^{\dagger}_{3L} \phi L_{3RKK}\Big].
\end{split}
\end{equation}
By integrating out the $Q_{3KK}$ and $L_{3KK}$ mediators we obtain the dominant effective Yukawa terms
\begin{equation}
\begin{split}
\mathcal{L}_{Y_{3j}}=&  \left(\frac{2}{R\Lambda\pi }\right)\frac{2}{\pi R\Lambda}  \left[\braket{\phi}({R^{-2}+M_{Q_3}^2})^{-1/2}Y^{33}_Q
\right]
\Big[y_u^{3j} Q_{3L}^\dagger H_u u_{jR}+y_d^{3j} Q_{3L}^\dagger H_d d_{jR}\Big]
\\
&+ \left(\frac{2}{R\Lambda\pi }\right) \frac{2}{\pi R\Lambda} \left[\braket{\phi} ({R^{-2}+M_{L_3}^2})^{-1/2}Y^{33}_L\right]
\Big[y_e^{3j} L_{3L}^\dagger H_d e_{jR}\Big]
\label{eq:effyuk3j}
\end{split}
\end{equation}
The above Yukawa matrices consist of only the third row elements being non-zero.
Ignoring the subdominant couplings below, 
they may be diagonalised by rotating
$u_{jR},d_{jR},e_{jR}$ to yield only non-zero (3,3) elements corresponding to the $t,b,\tau$ masses, respectively.
This explains why the third family is the heaviest one, which is due to $Q_{3KK}$ and $L_{3KK}$
being the lightest KK states.

To populate the more than just the (3,3) elements of the Yukawa matrices, we must consider the effect of the next lightest states, namely the rest of the third family 
$d_{3KK},u_{3KK},e_{3KK}$ states.
We first write down the interactions involving $d_{3KK},u_{3KK},e_{3KK}$, which is 
another subset of the terms in Eqs.\ref{L0Q}, \ref{L0L},
\begin{equation}
\begin{split}
\mathcal{L}_{q_{3KK}}&= 
\frac{2}{\pi R\Lambda}  \Big[ y_u^{i3} (Q_{iL}^\dagger H_u u_{3RKK}) +u\leftrightarrow d\Big]
+\left(\frac{2}{R\Lambda\pi }\right) \Big[ Y_u^{33} {u_{3LKK}}^{\dagger} \phi u_{3R}+u\leftrightarrow d\Big],
\end{split}
\end{equation}
\begin{equation}
\begin{split}
\mathcal{L}_{e_{3KK}}&= 
 \frac{2}{\pi R\Lambda}  \Big[ y_e^{i3} (L_{iL}^\dagger H_d e_{3RKK}) \Big]
+ \left(\frac{2}{R\Lambda\pi }\right) \Big[ Y_e^{33} {e_{3LKK}}^{\dagger} \phi  e_{3R}\Big].
\end{split}
\end{equation}
By integrating out the $d_{3KK},u_{3KK},e_{3KK}$ mediators we obtain the sub-dominant effective Yukawa terms
\begin{equation}
\begin{split}
\mathcal{L}_{Y_{i3}}=& \left(\frac{2}{R\Lambda\pi }\right)\frac{2}{\pi R\Lambda}  \left[\braket{\phi}({R^{-2}+M_{u_3}^2})^{-1/2}Y^{33}_u
\right ]\Big[y_u^{i3} Q_{iL}^\dagger H_u u_{3R}\Big] +u\leftrightarrow d
\\
&+ \left(\frac{2}{R\Lambda\pi }\right)\frac{2}{\pi R\Lambda} \left[\braket{\phi}({R^{-2}+M_{e_3}^2})^{-1/2}Y^{33}_e
\right]\Big[y_e^{i3} L_{iL}^\dagger H_d e_{3R}\Big]
\label{eq:effyuki3}
\end{split}
\end{equation}

The Yukawa matrices in LR convention, resulting from the third family KK mediators (assumed to be the lightest ones and the only ones considered so far),
now consist of the dominant third row elements from Eq.\ref{eq:effyuk3j}
and the sub-dominant third column elements from Eqs.\ref{eq:effyuki3}, i.e. only the third row and third column have non-zero elements.
In the basis where the dominant third row is diagonalised, the Yukawa matrices may be expressed in the form~\footnote{To arrive at this form, starting from Yukawa matrices with only non-zero third rows and third columns,
 we have first rotated $(u_{1R},u_{2R})$, $(d_{1R},d_{2R})$, $(e_{1R},e_{2R})$ to put zeroes in the first column,
then rotated $(Q_1,Q_2)$ to set $Y^{u}_{13}=0$, then finally we have rotated $(u_{2R},u_{3R})$, $(d_{2R},d_{3R})$, 
$(e_{2R},e_{3R})$ to diagonalise the dominant third row.}, ignoring overall factors and the compactification scale,
\bea
Y^{u}_{ij}&=&  \pmatr{0& 0 & 0 \\
0 & x^{u}_{22} & x^{u}_{23}\\
0 & x^{u}_{32} & x^{u}_{33}
}
\frac{Y^{33}_u\langle \phi \rangle }{M_{u_3}}
+\pmatr{
0 & 0 & 0 \\
0 & 0 & 0 \\
0 & 0 & y^{u}_{33}} 
\frac{Y^{33}_Q\langle \phi\rangle }{M_{Q_3}} \nonumber \\
Y^{d}_{ij}&= &  \pmatr{
0 &x^{d}_{12} & x^{d}_{13} \\
0 & x^{d}_{22}& x^{d}_{23}\\
0 & x^{d}_{32}& x^{d}_{33}
}
\frac{Y^{33}_d\langle \phi \rangle }{M_{d_3}}
+\pmatr{
0 & 0 & 0 \\
0 & 0 & 0 \\
0 & 0 &  y^{d}_{33}} 
\frac{Y^{33}_Q\langle \phi \rangle }{M_{Q_3}}\nonumber \\
Y^{e}_{ij}&= &  
 \pmatr{
0 &x^{e}_{12} & x^{e}_{13} \\
0 & x^{e}_{22}& x^{e}_{23}\\
0 & x^{e}_{32}& x^{e}_{33}
}
\frac{Y^{33}_e\langle \phi \rangle }{M_{e_3}}
+\pmatr{
0 & 0 & 0 \\
0 & 0 & 0 \\
0 & 0 & y^{e}_{33}} 
\frac{Y^{33}_L\langle \phi \rangle }{M_{L_3}}
\label{Yuk_mass_insertion_ud}
\eea
where the Yukawa couplings $x^{u,d,e}_{ij}$, $y^{u,d,e}_{33}$ are linear combinations of the original Yukawa couplings
to the third family KK modes $y^{u,d,e}_{i3}$, $y^{u,d,e}_{3j}$, and are hence 
expected to be of order unity. We also assume that they are defined to absorb all the prefactors, which are also of order unity.
The Yukawa hierarchy in this model is generated by the previously assumed mass hierarchy 
of 5d masses, 
$M_{Q_3}\ll M_{d_3} \ll M_{u_3}$, and $M_{L_3}\ll M_{e_3}$,
called ``messenger dominance'' in  \cite{Ferretti:2006df}. The scale of the actual masses must comply with Eq. \ref{eq:comp}.
This implies that the second matrices on the right-hand side of the equalities in Eq.\ref{Yuk_mass_insertion_ud}
dominate, giving the dominant third family masses $m_t$, $m_b$, $m_{\tau}$.
The sub-dominant first matrices on the right-hand side of the equalities in Eq.\ref{Yuk_mass_insertion_ud}
are responsible for the second family masses with \cite{Zyla:2020zbs}
\begin{equation}
\frac{m_{c}}{m_{t}}\sim \frac{M_{Q_3}}{M_{u_3}}\sim 0.0074,\ \ \ \frac{m_{s}}{m_{b}}\sim\frac{ M_{Q_3}}{M_{d_3}}\sim 0.023,
\end{equation}
with a more pronounced mass hierarchy
in the up sector than the down sector.
It also implies non-zero quark mixing angles, 
with small values of the CKM elements $|V_{ub}|\sim |V_{cb}|\sim m_s/m_b$, which is a successful prediction for $|V_{cb}|$.
However $|V_{ub}|$ is too large and 
the smallness of the Cabibbo angle is not explained with $|V_{us}|\sim 1$.
It also implies a natural lepton mass hierarchy
\begin{equation}
 \frac{m_{\mu}}{m_{\tau}} \sim \frac{M_{L_3}}{M_{e_3}}\sim 0.59,
 \end{equation} 
 with non-zero charged lepton contributions to leptonic mixing angles analogous to the quark ones. However the neutrino sector will dominantly contribute to leptonic mixing angles as discussed in the next section.
At this stage the first family quarks and leptons are massless, but will develop small masses when the first and/or second family KK mediators are included.

We remark that the generation of the hierarchies between 5d quark masses not only implies a hierarchy between the actual quark masses, but also implies small mixing angles. Therefore the assumption is predictive and not only a reparametrization.

\section{Neutrino Majorana masses from KK Dirac neutrinos}
\label{neutrino}

\begin{figure}[h]
\centering
\includegraphics[scale=0.4]{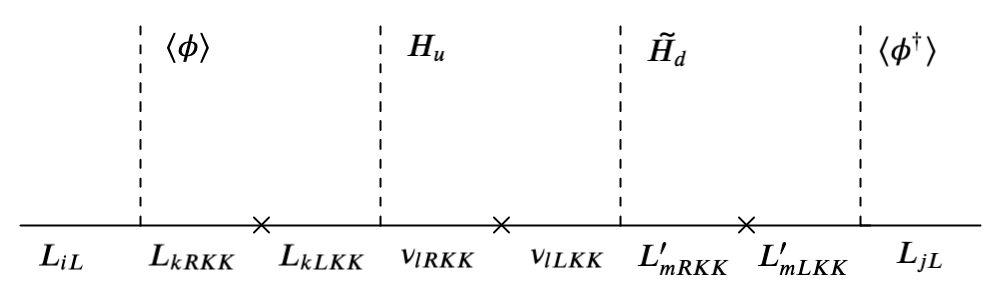}
		\caption{Diagrams for the type Ib seesaw mechanism for neutrino masses.}
		\label{fig:YF2}
\end{figure}

Finally, to generate the neutrino masses, we follow a somewhat different procedure to the case of charged lepton and quark masses.
The Higgs doublets do not couple to the massless chiral lepton doublets $L_i$ to the neutrino singlets $\nu$, as seen in Eq.\ref{Yuks}, for the same reason that SM Yukawa couplings are not allowed, namely that the Higgs carry $U(1)'$ charge while the SM chiral fermions do not. In the case of charged lepton and quark masses, such Yukawa couplings are mediated by $U(1)'$ charged KK excitations, arising from the $SO(3)$ lepton and quark triplets.
However, in the case of neutrinos, the 5d neutrino singlets $\nu$ are chosen to be $SO(3)$ singlets, and do not have any zero modes,
and therefore the KK excitations are neutral under $U(1)'$ and so do not couple to $L_i$.
In this case the KK mediation arises from the following couplings,
\begin{equation}
\begin{split}
\mathcal{L}_{0\nu}&= 
\frac{2}{\pi R\Lambda} \Big[ y_\nu^i L_{iLKK}^\dagger H_{u} \nu_{RKK}+y^{'j}_\nu L^{'}_{jRKK} \tilde{H}_{d} \nu^\dagger_{LKK}\Big],
\end{split}
\end{equation}
where $\tilde{H}_d$ is the CP conjugate of the Higgs doublet $H_d$, 
and hence $\tilde{H}_d$ will have the same hypercharge as $H_u$ but opposite
isocharge. As mentioned before, the $\mathbb{Z}_3$ symmetry enforces the $H_{u,d}$ couplings corresponding to up-type and down-type fermions respectively, but allows the Neutrino Majorana mass terms.
The $L_{iKK}$ and $L'_{iKK}$ have opposite $U(1)'$ charges but both will couple to the neutral massless chiral lepton doublet $L_i$. 
It is worth to note that these couplings involve two KK excitations in each term.
Also note that two different Higgs doublets are involved here, which couple to the two different KK lepton doublets. 
The two different KK neutrino fields above, $\nu_{LKK}$ and $\nu^c_{RKK}$, form a single heavy Dirac KK mass, which 
may be integrated out to generate the effective neutrino operators,
\begin{equation}
\mathcal{L}_{0\nu}= \frac{64 R\Lambda }{\pi^5 \Lambda^3}\braket{\phi\phi^\dagger} y^k_\nu y^{'l}_\nu Y_L^{ki}Y_L^{lj} L_{iL}^\dagger H_u \tilde{H}_d L_{jL}.
\label{Wb}
\end{equation}
Note that this is not the usual Weinberg operator since it involves two different Higgs doublets, so it is a new type of Weinberg
operator, originating from a variant of the type I see saw mechanism called the type Ib seesaw mechanism~\cite{Hernandez-Garcia:2019uof}. 
The new effective dimension 5 operator is mediated by the Dirac KK neutrino modes, as shown in figure \ref{fig:YF2}. In fact this version of the type Ib seesaw mechanism is slightly different to that proposed in~\cite{Hernandez-Garcia:2019uof}, since it involves two additional scalar singlets $\phi\phi^\dagger$, leading to additional mass suppression, but the basic features are the same: two different Higgs doublets with a single Dirac heavy neutrino mass mediating the diagram, leading to light effective Majorana neutrino masses.
However, assuming zero 5d Dirac masses,
the new mechanism does not so far explain the smallness of neutrino mass unless the compactification scale is quite high,
or at least one of the Yukawa couplings is very small.

Now including also the explicit 5d Dirac masses discussed in Eq.\ref{eq:5m} of 
the previous section, the operator in Eq.~\ref{Wb} is generalised to 
\begin{equation}
\begin{split}
\mathcal{L}_{0\nu}=&\frac{32}{\pi^5 R^2\Lambda^2} \braket{\phi\phi^\dagger} \ y^k_\nu y^{'l}_\nu Y^{ki}_L Y^{lj}_L L_{iL}^\dagger H_u \tilde{H}_d L_{jL}\times\\
& \big(R^{-2}+(M_\nu)^2\big)^{-1/2}\big(R^{-2}+(M_L^i)^2\big)^{-1/2}\big(R^{-2}+(M_{L}^j)^2\big)^{-1/2}.
\end{split}
\end{equation}
This leads to a general symmetric Majorana mass matrix for the neutrinos from a single pair of 5d Dirac neutrinos.
There is now the opportunity to explain the smallness of neutrino mass even with a low compactification scale and 
Yukawa couplings of order unity, by assuming that one of the 5d Dirac masses is very large. For example the 
($SO(3)$ singlet) 5d neutrino mass $M_\nu$
could be much larger than the ($SO(3)$ triplet) 5d lepton doublet 
masses, $M_\nu \gg M_L^i$, leading to highly suppressed neutrino masses,
as in the traditional type Ia seesaw mechanism. One may speculate on mechanisms which would result
in a larger mass for the $SO(3)$ singlet than the $SO(3)$ triplet fermions, but that is beyond the scope of the present paper.

\section{Non universal $Z'$ couplings and phenomenology}
\label{pheno}

\begin{figure}
 \centering
\includegraphics[scale=0.4]{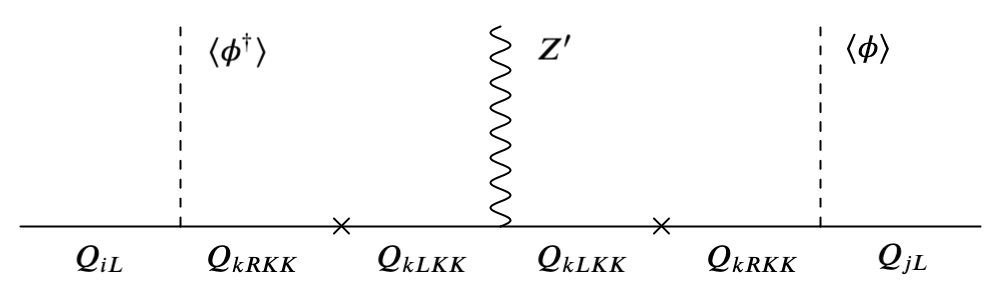}
		\caption{Diagrams for the effective $Z'$ coupling to the isoneutral quark doublets $Q_{iL}$, 
		mediated by the isocharged KK excitations (where  $Z'$ only couples to isocharged states).
		Similar diagrams may be drawn for 
		all the SM fermions $Q_{iL},u_{iR},d_{iR},L_{iL},e_{iR}$, which are isoneutral, with effective $Z'$ couplings
		mediated by their respective isocharged KK excitations.}
		\label{fig:ZZ}
\end{figure}

The terms that would generate an effective 4d $Z'$ coupling would be \cite{King:2018fcg}
\begin{equation}
\mathcal{L}_{Z'}=\left(\frac{2}{R\Lambda\pi}\right)Y^{ij}_\psi \psi^\dagger_{i} \phi\psi_{jKK}+g\left(\frac{2}{\pi}\right)^{3/2}\psi_{iKK}^\dagger \gamma^\mu Z'_\mu \psi_{iKK}+h.c.,
\end{equation}
where can be any  $\psi=Q_L,u_R,d_R,L_L,e_R$.  By integrating out the KK modes, as in fig. \ref{fig:ZZ}, we obtain the non universal corrections
\begin{equation}
\mathcal{L}_{Z'}=g\left(\frac{2}{R\Lambda\pi}\right)\left[\braket{\phi\phi^\dagger}({R^{-2}+M_{\psi\ i}^{2}})^{-1}
\right]Y^{ij}_\psi\psi_{i}^\dagger \gamma^\mu Z'_\mu \psi_{j}.
\end{equation}

These $Z'$ couplings are not universal since we are assuming $M_{Q_3}\ll M_{d_3} \ll M_{u_3}$, and $M_{L_3}\ll M_{e_3}$, 
in order to generate the desired Yukawa hierarchies. Indeed the $Z'$ couplings are related to the Yukawa couplings as in \cite{King:2018fcg}.
The dominant $Z'$ couplings will be generated by the lightest messenger masses associated with the third family doublets $Q_3$ and $L_3$,
\begin{equation}
\begin{split}
& \mathcal{L}_{Z'Q_3}= g\left(\frac{2}{R\Lambda\pi}\right)\left[\braket{\phi\phi^\dagger}({R^{-2}+M_{Q_3}^{2}})^{-1}\right]Y^{33}_QQ_{3L}^\dagger \gamma^\mu Z'_\mu Q_{3L}
\\
&
\mathcal{L}_{Z'L_3}=g\left(\frac{2}{R\Lambda \pi}\right)\left[\braket{\phi\phi^\dagger}({R^{-2}+M_{L_3}^{2}})^{-1}\right]Y^{33}_L L_{3L}^\dagger \gamma^\mu Z'_\mu L_{3L}.
\label{Zp3}
\end{split}
\end{equation}
Note that the prefactors in Eq.~\ref{Zp3} are related to the square of the prefactors of the dominant Yukawa couplings in 
Eq.~\ref{eq:effyuk3j}.
Thus, this model not only explains the origin of the Yukawa couplings but also relates them to non universal $Z'$ couplings \cite{King:2018fcg}. Unlike \cite{King:2018fcg},
this model is enhanced with extra dimensions, which explains the origin of the mediators and the $U(1)'$ breaking field $\phi$.

\begin{figure}
\begin{minipage}{0.32\linewidth}
\centerline{\includegraphics[width=0.9\linewidth]{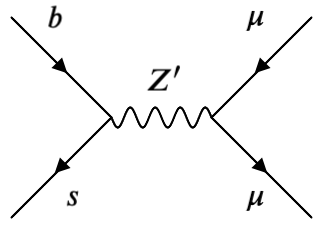}}
\end{minipage}
\begin{minipage}{0.32\linewidth}
\centerline{\includegraphics[width=0.9\linewidth=true]{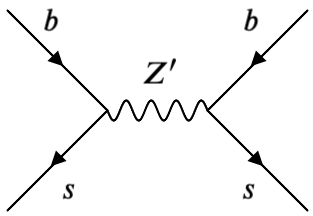}}
\end{minipage}
\begin{minipage}{0.32\linewidth}
\centerline{\includegraphics[width=0.9\linewidth]{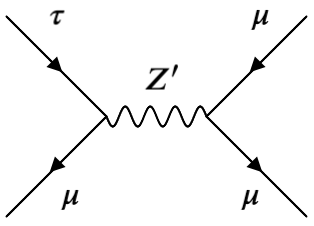}}
\end{minipage}
\caption{These $Z'$ exchange diagrams 
contribute to $R_{K^{(^*)}}$ (left), to $B_s$ mixing (centre)
and to $\tau\rightarrow \mu \mu \mu $ (right).
The couplings are defined as $g_{bs}Z'_{\mu}b_L^{\dagger}\gamma^{\mu}s_L$,
$g_{\mu \mu}Z'_{\mu}\mu_L^{\dagger}\gamma^{\mu}\mu_L$ and $g_{\tau \mu}Z'_{\mu}\tau_L^{\dagger}\gamma^{\mu}\mu_L$.}
\label{pheno2}
\end{figure}

The non universal $Z'$ couplings in Eq.~\ref{Zp3}
may help to generate the non universal leptonic decays \cite{Aaij:2014ora} which can't be explained within the SM \cite{Descotes-Genon:2013wba,Altmannshofer:2013foa,Ghosh:2014awa}.
Although the chiral fermions do not carry $U(1)'$ charges, the diagrams in Fig.~\ref{fig:ZZ}
generate effective $Z'$ couplings to chiral fermions, via the KK messengers do carry $U(1)'$ charges
(which are trivially anomaly free).
The $Z'$ couplings in the above basis are dominated by left-handed couplings
to the third family in Eq.~\ref{Zp3}, which we write approximately as~\cite{King:2019hkt},
\be
y_t^2 g Z'_{\mu}Q^{\dagger}_{3L}\gamma^{\mu}Q_{3L}+
y_{\tau}^2 g Z'_{\mu}L^{\dagger}_{3L}\gamma^{\mu}L_{3L},
\label{Zp_Q3L3}
\ee
to emphasise the approximate relation to the top and tau Yukawa couplings, $y_t$ and $y_{\tau}$,
resulting from Eq.~\ref{eq:effyuk3j}.
Flavour changing couplings involving the quark doublets 
$Q_{3L}=(t, b)^T_L$, $Q_{2L}=(c, s)^T_L$, will be generated
when the Yukawa matrices in Eq.~\ref{Yuk_mass_insertion_ud} are diagonalised,
 \be
y_t^2 g Z'_{\mu}Q^{\dagger}_{3L}\gamma^{\mu}Q_{3L} \ 
\rightarrow \ 
V_{ts}Z'_{\mu}Q^{\dagger}_{3L}\gamma^{\mu}Q_{2L},\ \ 
V_{ts}^2Z'_{\mu}Q^{\dagger}_{2L}\gamma^{\mu}Q_{2L}, \ \ldots \ 
\rightarrow \ 
V_{ts}Z'_{\mu}b_L^{\dagger}\gamma^{\mu}s_L,  \ \ldots
\label{Zp_Q_couplings_ij}
\ee
Similarly the operator $y_{\tau}^2 g Z'_{\mu}L^{\dagger}_{L3}\gamma^{\mu}L_{L3}$ 
in Eq.\ref{Zp_Q3L3} leads to 
flavour changing couplings involving the
lepton doublets $L_{3L}=(\nu_{\tau}, \tau )^T_L$, $L_{2L}=(\nu_{\mu}, \mu )^T_L$,
controlled by a left-handed lepton mixing $\theta_{23}^e$,
\be
\! \! \! \! 
\theta_{23}^ey_{\tau}^2 Z'_{\mu}L^{\dagger}_{3L}\gamma^{\mu}L_{2L},\ \ 
(\theta_{23}^e)^2y_{\tau}^2 Z'_{\mu}L^{\dagger}_{2L}\gamma^{\mu}L_{2L} 
  \ \ldots \ 
\rightarrow \ \ \ 
\theta_{23}^ey_{\tau}^2 Z'_{\mu}\tau_L^{\dagger}\gamma^{\mu}\mu_L,\ \ 
(\theta_{23}^e)^2y_{\tau}^2 Z'_{\mu}\mu_L^{\dagger}\gamma^{\mu}\mu_L  
\label{Zp_L_couplings_ij}
\ee
where we have taken $y_t\approx g\approx 1$. 
The couplings in Eqs.\ref{Zp_Q_couplings_ij}, \ref{Zp_L_couplings_ij}
control the $Z'$ exchange diagrams in Fig.\ref{pheno2} which 
contribute to $R_{K^{(^*)}}$ (left), to $B_s$ mixing (centre)
and to $\tau\rightarrow \mu \mu \mu $ (right).

The $Z'$ contributes to $R_{K^{(^*)}}$ at tree-level, 
via the (left) diagram in Fig.\ref{pheno2},
where the requirement to explain the anomaly is, 
defining the couplings of the $Z'$ to the relevant fermions as $g_{\mu \mu}$ and $g_{bs}$,
\be
\label{RKuirement2}
\frac{g_{\mu \mu} g_{bs} }{M_{Z'}^2}\approx \frac{y_{\tau}^2  (\theta_{23}^e)^2 V_{ts}}{M_{Z'}^2} \approx
\frac{1.1}{(35~\rm{TeV})^2} 
%\rightarrow 
%y_{\tau}^2  (\theta_{23}^e)^2 \approx
%2.2\times 10^{-2}  \left( \frac{M_{Z'}}{1~\rm{TeV}} \right)^2,
\ee
Since $V_{ts} \approx 4.0\times 10^{-2}$,
this requires quite a large $y_{\tau}\approx 1$ 
(i.e. large $\tan \beta =\langle H_u \rangle / \langle H_d \rangle  $) and a large mixing angle $\theta_{23}^e \approx 0.1$,
together with a low mass $M_{Z'}\approx 1$ TeV, close to current LHC limits~\cite{Falkowski:2018dsl}.

%%%%%%%%%%%

Now $B_s$ mixing is mediated by tree-level $Z'$ exchange as in the (centre) diagram in Fig.\ref{pheno2},
leading to the 2015 bound~\cite{Falkowski:2018dsl},
\be
\frac{g_{bs} g_{bs} }{M_{Z'}^2}
 \approx 
  \frac{ V_{ts}^2  }{M_{Z'}^2}
 \leq \frac{1}{(140~\rm{TeV})^2}
 % \rightarrow 
%M_{Z'}\geq V_{ts}(140~\rm{TeV})= 5.6~\rm{TeV}
\ee
leading to $M_{Z'}\geq 5.6~\rm{TeV}$, in some tension with the $R_{K^{(^*)}}$ requirement above.
However the stronger 2017 bound with scale of $770$ TeV instead of $140$ TeV implies a bound of 
$M_{Z'}\geq 31~\rm{TeV}$, which is quite incompatible with the $R_{K^{(^*)}}$ requirement in Eq.\ref{RKuirement2}.

The compactification scale has an experimental lower bound from Eq. \ref{eq:comp}. The $Z'$ mass given by the VEV $\braket{\phi}$ is smaller than the compactification scale, due to the assumed radiative breaking mechanism, so that the $Z'$ effects are larger than the SM KK modes. Pushing up the $Z'$ mass also pushes up the compactification scale.

Moreover $\tau\rightarrow \mu \mu \mu $ is mediated by tree-level $Z'$ exchange as in the (right) diagram in Fig.\ref{pheno2},
leading to the bound~\cite{Falkowski:2018dsl},
\be
\frac{g_{\tau \mu} g_{\mu \mu} }{M_{Z'}^2}
 \approx 
  \frac{(\theta_{23}^e)^3y_{\tau}^4 }{M_{Z'}^2}
 \leq \frac{1}{(16~\rm{TeV})^2}
% \rightarrow 
%y_{\tau}^4  (\theta_{23}^e)^3 \leq
%4.0\times 10^{-3}  \left( \frac{M_{Z'}}{1~\rm{TeV}} \right)^2.
\ee

Writing $g_{\tau \mu} = g_{\mu \mu}/\theta^e_{23}$,
the bounds on $B_s$ mixing and $\tau\rightarrow \mu \mu \mu $ may be written as:
\be
\frac{g_{bs}}{M_{Z'}}
 \leq \frac{1}{(140~\rm{TeV})}, \ \ \ \ 
 \frac{g_{\mu \mu} }{M_{Z'}}
  \leq \frac{(\theta^e_{23})^{1/2}}{(16~\rm{TeV})}
\label{bound1}
\ee
which may be combined, leading to a bound~\footnote{I am grateful to E.Perdomo for pointing out this bound.} on the contribution to $R_{K^{(^*)}}$~\cite{King:2019hkt,King:2020fdl},
\be
\frac{g_{\mu \mu} }{M_{Z'}} \frac{g_{bs}}{M_{Z'}}
 \leq \frac{(\theta^e_{23})^{1/2}}{(140~\rm{TeV})(16~\rm{TeV})}
 =\frac{(\theta^e_{23})^{1/2}}{(47~\rm{TeV})^2}
 \label{combined_bound}
 \ee
 which is somewhat less than the $\frac{1.1}{(35~\rm{TeV})^2}$ 
 required in Eq.\ref{RKuirement2} to explain the anomaly.
Moreover, the stronger 2017 bound with scale of $770$ TeV instead of $140$ TeV implies a bound of 
$\frac{(\theta^e_{23})^{1/2}}{(111~\rm{TeV})^2}$, which is significantly less than
the $\frac{1.1}{(35~\rm{TeV})^2}$ required to explain the anomaly. 

The $R_{K^{(^*)}}$ anomaly could be explained in this model, consistently with $\tau\rightarrow \mu \mu \mu $, if the 
second lepton doublet messenger mass were assumed to be of the same order as that of the third family, namely 
$M_{L_2}\sim M_{L_3}\ll M_{e_3}$, which would imply that a new $Z'$ coupling to muons $g_{\mu \mu}$
could arise independently of the $\tau$ coupling $g_{\tau \tau}$, removing the bound on $g_{\mu \mu}$ in Eq.\ref{bound1}.
In this case the bound in Eq.~\ref{combined_bound} would be evaded, and indeed 
$\theta^e_{23}$ could be naturally very small,
as expected, say of order $V_{ts}$. 
However, in such a scenario, the smallness of the muon mass compared 
to the $\tau$ mass could not 
be explained by the hierarchy of 5d Dirac masses, but instead would have to be accounted for by a tuning of Higgs Yukawa
couplings, as in the Standard Model. Therefore, we would prefer that the $R_{K^{(^*)}}$ rate is much closer to the Standard Model
prediction, so that we can provide a natural explanation of the smallness of the muon mass as being due to 
$M_{L_3}\ll M_{e_3},M_{L_2}$. In this preferred scenario, although we cannot explain the current $R_{K^{(^*)}}$ anomaly,
nevertheless there can be non-zero contributions to all the three BSM processes in Fig.\ref{pheno2}, which could be all observed 
in the future.

\section{Conclusion}
\label{conclusion}

In this paper we have proposed a theory of quark and lepton masses with non-universal $Z'$ couplings based on 
a simple extension to the Standard Model in which quarks and leptons are promoted to 5d 
gauged $SO(3)$ isospin triplets. 
We emphasise that this is not weak isospin, nor is it a family symmetry, it is a completely new degree of freedom carried by each 5d multiplet $Q^{\alpha}_i,u^{\alpha}_i,d^{\alpha}_i,L^{\alpha}_i,e^{\alpha}_i$, where $i=1,2,3$ is a family index and $\alpha=1,2,3$ is a new
$SO(3)$ index.
In the 4d effective theory, the $SO(3)$ is broken to $U(1)'$, 
under which the triplets carry isocharges $(+1,0,-1)$.
The breaking is achieved via the 
$S^1/(\mathbb{Z}_2\times \mathbb{Z}'_2)$ orbifold, with $U(1)'$ subsequently spontaneously
broken, resulting in a massive $Z'$.

Quarks and leptons in the 5d bulk appear as massless modes, isoneutral under $U(1)'$.
The Higgs doublets are located on the brane and have isocharge $\pm 1$ (ignoring the heavier isoneutral Higgs doublets).
There are zero Yukawa 
couplings to the Higgs, and zero couplings to $Z'$, due to the $U(1)'$ symmetry.
However, after the $U(1)'$ breaking, both 
Yukawa couplings and non-universal $Z'$ couplings are generated by heavy Kaluza-Klein exchanges.
This may be regarded as the ultraviolet completion of a model proposed some time ago based on a vector-like fourth family
isocharged under a $U(1)'$, which mediates both Yukawa couplings and $Z'$ couplings for the chiral quarks and leptons,
which are isoneutral under $U(1)'$, thereby relating such couplings. 
The idea here is that the fourth vector-like family is identified as a KK excitation along an extra 5th compact
dimension. 

However, such a KK interpretation required the introduction of the $SO(3)$
isospin, broken to $U(1)'$, to enable the combination of isoneutral chiral fermions and isocharged KK excitations.
The presence of a brane scalar
that breaks the
$U(1)'$ originates provides a satisfying mechanism for symmetry breaking.
The hierarchical Yukawa couplings of charged fermions results from a hierarchy of 5d Dirac fermion masses, in particular,
the lightest masses being associated with the third family quarks and leptons, with 
$1/R \lesssim  M_{Q_3}\ll M_{d_3} \ll M_{u_3}$, and $1/R \lesssim M_{L_3}\ll M_{e_3}$, where the two other families are assumed to have even heavier masses.
Majorana neutrino mass and mixing arises from a novel type Ib seesaw mechanism, mediated by Kaluza-Klein Dirac neutrinos
with large 5d Dirac masses $M_\nu \gg M_L^i$.
This mechanism could be applied to any other extra dimensional and/or string model, in order to obtain Majorana neutrinos from a seesaw mechanism in which the mediators are Kaluza-Klein Dirac neutrinos.

In the present model, since each quark and lepton field forms a complete isospin triplet,
$Q^{\alpha}_i,u^{\alpha}_i,d^{\alpha}_i,L^{\alpha}_i,e^{\alpha}_i$, each field will have three
isocharges $(+1,0,-1)$, with only the isoneutral ones having massless zero modes.
However, for each flavour and isospin index $(i,\alpha)$, there will be 
an infinite KK tower of massive Dirac (or vector-like) 
states, providing a wealth of new states which could be discovered at future colliders, although given that 
the 5d masses of these states are very hierarchical, only the lightest ones above will
be discovered to start with.
Thus the lightest KK modes are the electroweak doublets and isotriplets
$Q^{\alpha}_{3KK}=(T,B)^{\alpha}$ with mass $M_{Q_3}$
and $L^{\alpha}_{3KK}=(N,E)^{\alpha}$ with mass $M_{L_3}$, which 
automatically respect a custodial $SU(2)$ symmetry, allowing the compactification scale to be as low as the direct collider limits
on universal extra dimensions, around the TeV scale~\cite{Deutschmann:2017bth}.
In the present model there also be an isotriplet of Higgs doublets $H_{u,d}^{\alpha}$ on the zero brane plus 
an additional pair of isoneutral Higgs doublets $H_{u,d}^{0'}$ on the other brane, all of which could be within experimental reach.

The non-universal  $Z'$ couplings may contribute to 
semi-leptonic $B$ decay ratios $R_{K^{(*)}}$ which violate $\mu - e$ universality, which in this model 
are related to the origin of the fermion Yukawa couplings. 
However, the natural expectation is that the presently indicated rate of $R_{K^{(*)}}$ is too large to be
explained in this model, although it 
could be observed at a lower rate in future, along with other BSM signals of $B_s$ mixing 
and $\tau\rightarrow \mu \mu \mu $.

In conclusion, we have extended the SM fermions into a flat 5d bulk, with Higgs doublets on the branes, in order to shed light on the origin of Yukawa couplings.
We were led to introduce a new $SO(3)$ isospin under which fermions and Higgs are isotriplets, although extra isosinglet Higgs and isosinglet neutrinos were also introduced.
We have assumed hierarchical 5d Dirac masses in order to account for the fermion mass hierarchies. However this is more than just a one-one repameterisation of the fermion masses by the 5d Dirac masses. With the simple assumptions $1/R \lesssim  M_{Q_3}\ll M_{d_3} \ll M_{u_3}$, and $1/R \lesssim M_{L_3}\ll M_{e_3}$ we have reproduced all the charged fermion mass hierarchies and generated small quark mixing predominantly from the down quark sector. 
Majorana neutrino masses arise from Dirac KK neutrino exchange, via a type Ib seesaw mechanism,
with $M_\nu \gg M_L^i$ accounting for the smallness of neutrino mass.

Finally we note that, whilst similar results could also be achieved in 4d by adding a fourth vector-like family to the SM, we have shown that by introducing an extra dimension and 5d isospin $SO(3)$ we can explain the origin of the fourth vector-like family (and the fifth one which is necessary to obtain non-zero first family quark and lepton masses) by the very simple and elegant extension to the SM 
shown in Table~\ref{ta:fcq}. Thus, the extra vector-like families are not introduced in an {\it ad hoc} way but instead emerge as KK excitations. 
Moreover, the experimental implications of such a novel theory of flavour
are very distinctive, being distinguishable from universal extra dimensions due to: the presence of the gauged $SO(3)$, 
broken to $U(1)'$ and leading to the $Z'$; the characteristic 5d Dirac mass pattern above;
and the rich spectrum of Higgs doublets on the branes. It would clearly be interesting to explore the phenomenology of this model further.

\subsection*{Acknowledgements}

S.\,F.\,K. acknowledges the STFC Consolidated Grant ST/L000296/1 and the European Union's Horizon 2020 Research and Innovation programme under Marie Sk\l{}odowska-Curie grant agreements Elusives ITN No.\ 674896 and InvisiblesPlus RISE No.\ 690575. 

\appendix

\section{$S^1/(\mathbb{Z}_2\times \mathbb{Z}'_2)$ orbifold}
\label{app:orb}

We assume a 5d spacetime, with an extra spatial dimension $y$. The extra dimension is compactified as an orbifolded circle with the identifications
\begin{equation}
\begin{split}
y&\sim y+2\pi R,\\
y&\sim -y,
\label{eq:orby}
\end{split}
\end{equation}
which defines the orbifold geometry.
The orbifolding leaves 2 fixed points
\begin{equation}
\bar{y}=\{0,\pi R/2\},
\end{equation}
which allow boundary conditions on the fields. Any arbitrary field $\Phi(x,y)$ must comply with eq. \ref{eq:orby}, up to a gauge transformation, which defines the boundary conditions
since
\begin{equation}
\begin{split}
\Phi(x,y)&=P_0\ \Phi(x,-y),\\
\Phi(x,y+\pi R/2)&=P_{\pi R/2}\ \Phi(x,-y+\pi R/2)
\end{split}
\end{equation}
so that the two independent boundary conditions satisfy $P_{0,\pi R/2}^2=1$ and they belong to the extended gauge group. Each condition corresponds to each $\mathbb{Z}_2$ \cite{Hebecker:2001jb}.
 
We locate all fields in the bulk which have to belong to irreducible representations of the Lorentz group $SO(1,4)$. It is important to understand how each field transforms under the orbifolded parity.

The orbifold operation is the 5th parity operator $\mathcal{P}_5$ which is accompanied by a gauge transformation $P_{0,\pi R}$ depending on which fixed point the parity is applied.

The simplest field field is a 5d scalar $\phi(x,y)$ which transforms as
\begin{equation}
\begin{split}
\phi(x,y)&=\tilde{\mathcal{P}}_5\phi(x,y)=P\mathcal{P}_5\phi(x,y)=P_{0}\ \phi(x,-y),\\
\phi(x,\pi R/2+y)&=\tilde{\mathcal{P}}_5\phi(x,\pi R/2+y)=P_{\pi R/2}\mathcal{P}_5\phi(x,\pi R/2+y)=P_{\pi R/2}\ \phi(x,\pi R/2-y),\\
\end{split}
\end{equation}
where $0,\pi R/2$ are the fixed branes. After compactification, the 5d scalar becomes a 4d scalar.

One can also have vector fields $A_M(x,y)$ where $M=0,1,2,3,5$. The vector transforms as
\begin{equation}
\begin{split}
A_M(x,y) =P_0\mathcal{P}_5 A_M(x,y) \to & P_0\mathcal{P}_5 A_\mu(x,y)=P_{0}A_\mu(x,
-y),\\
\to & P_0\mathcal{P}_5 A_5(x,
y)=-P_{0}A_5(x,-y),\\
A_M(x,\pi R/2+y) =P_{\pi R/2}\mathcal{P}_5 A_M(x,\pi R/2+y) \to & P_{\pi R/2}\mathcal{P}_5 A_\mu(x,\pi R/2+y)\\ &\quad \quad=P_{\pi R/2}A_\mu(x,\pi R/2-y),\\
\to & \mathcal{P}_5 A_5(x,\pi R/2+y)=-P_{\pi R/2}A_5(x,\pi R/2-y),\\
\end{split}
\end{equation}
where the 5th component of the 5d vector field obtains an extra minus sign. After compactification the 5d vector decomposes into a 4d vector and a 4d scalar (the fifth component). 

The final representation is a 5d spinor $\Psi (x,y)$. For 5d spinors one has to enlarge the Clifford algebra. We can use the 4d Dirac matrices in the Weyl basis
\begin{equation}
\gamma^0=\left(\begin{array}{cc} 0 & I_2 \\ I_2 &0\end{array}\right), \ \ \ \gamma^i=\left(\begin{array}{cc} 0 & \sigma^i \\ -\sigma^i &0\end{array}\right),\ \ \ \gamma^5=\left(\begin{array}{cc} -I_2 & 0 \\ 0 & I_2\end{array}\right),
\end{equation}
where the $\gamma^5$ is the usual 4d one that is related to chirality.
Note that one can choose these as gamma matrices since they fulfill
\begin{equation}
\{\gamma^M, \gamma^N\}=2\eta^{MN}\mathbb{I}_4.
\end{equation}
The 5d spinor has 4 components which can be written in 4d terms as a Dirac fermion
\begin{equation}
\Psi(x,y)=\left(\begin{array}{c} \psi_R \\ \psi_L \end{array}\right),
\end{equation}
 where each $\psi_{L,R}$ is a Weyl fermion. The dynamical term $i\bar{\Psi}\gamma^M\partial_M \Psi$ mixes both $\psi_{L,R}$ in any basis, so in 5d the Dirac fermion is irreducible, i.e. a 5d fermion is a pair $\psi_{L,R}$ of Weyl fermions. The parity $\mathcal{P}_5$ operator for fermions involves a gamma matrix so
\begin{equation}
\begin{split}
\Psi(x,y)=P_0\mathcal{P}_5 \Psi(x,y)=P_{0}\ \gamma^5\Psi(x,-y)\to &-P_{0}\ \psi_R(x,-y),\\
& P_{0}\ \psi_L(x,-y),\\
\Psi(x,\pi R/2+y)=P_{\pi R}\mathcal{P}_5 \Psi(x,\pi R/2+y)\\
=P_{\pi R/2}\ \gamma^5\Psi(x,\pi R/2-y)\to &-P_{\pi R/2}\ \psi_R(x,\pi R/2-y),\\
& P_{\pi R/2}\ \psi_L(x,\pi R/2-y).
\end{split}
\end{equation}
After compactification, the 5d fermion decomposes into a $L,R$ 4d Weyl fermion pair. 

\section{$SO(3)$ gauge theory}
\label{app:so}

The $SO(3)$ group is the group of $3\times 3 $ orthogonal matrices with determinant of one. It has order 3, rank 1, and its generators are
\begin{equation}
T_{1}=\frac{1}{\sqrt{2}}\left(\begin{array}{ccc}0&1&0\\ 1 &0&1 \\
0&1&0\end{array}\right),\ \ \ T_{2}=\frac{1}{\sqrt{2}}\left(\begin{array}{ccc}0&-i&0\\ i &0&-i \\
0&i&0\end{array}\right),\ \ \ T_{3}=\left(\begin{array}{ccc}1&0&0\\ 0 &0&0 \\
0&0&-1\end{array}\right),
\end{equation}
where an element of the group can be written as
\begin{equation}
g=e^{i(\alpha T_{1}+\beta T_{2}+\gamma T_{3})}.
\end{equation}

Most fields $\Phi$ in our model will be in the adjoint representation $\Phi\sim(\textbf{3})$. One can write the 3 components of the field as
\begin{equation}
\Phi=\Phi_{1} T_{1}+\Phi_{2} T_{2}+\Phi_{3} T_{3},
\end{equation}
which transforms as
\begin{equation}
\Phi\to g\Phi g^{-1}.
\end{equation}
However it is much easier to write them in vector form as
\begin{equation}
\Phi=\left(\begin{array}{c} \Phi_1 \\ \Phi_2 \\ \Phi_3\end{array}\right),\ \ \ {\rm with \ transformation}\ \ \ \Phi\to g \Phi.
\end{equation}

We will impose the boundary condition
\begin{equation}
P=e^{i\pi T_{3}}=\left(\begin{array}{ccc}-1&0&0\\ 0&1&0 \\
0&0&-1\end{array}\right),
\end{equation}
which breaks $SO(3)\to SO(2)\simeq U(1)'$, where the $U(1)'$ is generated by $T_{3}$. 

The triplet field transforms as
\begin{equation}
\Phi\to P \Phi =\left(\begin{array}{c} -\Phi_1 \\ \Phi_2 \\ -\Phi_3\end{array}\right),
\end{equation}
which separates into an $SO(2)$ doublet or a charged pair and a neutral field.
A general $U(1)'$ gauge transformation would be
\begin{equation}
e^{i\alpha(x) T_3}=\left(\begin{array}{ccc}e^{i\alpha(x)}&0&0\\ 0&1&0 \\
0&0&e^{-i\alpha(x)}\end{array}\right),
\end{equation}
so that we can name the eigenstates by their charge
\begin{equation}
\Phi^0=\Phi_2,\ \ \ \Phi^+=\Phi_1, \ \ \ \Phi^-=\Phi_2.
\end{equation}
After the breaking $SO(3)\to U(1)'$, the field decomposes as
\begin{equation}
(\textbf{3})\to (0)+(1)+(-1).
\end{equation}
Under the $P$ transformation, the components of the $SO(3)$ triplet decompose are eigenstates with different eigenvalues. The neutral component have an eigenvalue $+1$ while the charged states have a $-1$ eigenvalue.

\section{Higgs localization and masses}
\label{app:h}

We assume that the physical Higgs is located in the zero brane. However we have assumed that it has a charge $-1$ under the $U(1)'$, but the zero brane has an unbroken $SO(3)$. To justify this, we locate the full triplet with components $H^+,H^0,H^-$ on the zero brane and further two copies in the $\pi R/2$ brane $H^{'-},H^{'0}$, where the $SO(3)$ is broken, as in table \ref{ta:hh}. 
As discussed below, the $H^-_{u,d}$ become the physical Higgses and are renamed as $H_{u,d}$ in the main text.

\begin{table}
\centering
\begin{tabular}{c|ccccc|c   }
 4d field &$SU(3)_C$ & $SU(2)_L$ & $U(1)_Y$ & $U(1)'$ & $\mathbb{Z}_3$  &  Brane\\
\hline
$H_u^+$ & $\textbf{1}$ &$\textbf{2}$ & $-1/2$&$+1$  &$\omega^2$& $0$   \\
$H_u^0$ & $\textbf{1}$ &$\textbf{2}$ & $-1/2$&$0$  &$\omega^2$& $0$   \\
$H_u^-$ & $\textbf{1}$ &$\textbf{2}$ & $-1/2$&$-1$  &$\omega^2$& $0$   \\
\hline
$H_d^+$ & $\textbf{1}$ &$\textbf{2}$ & $1/2$&$+1$  &$\omega^2$& $0$   \\
  $H_d^0$ & $\textbf{1}$ &$\textbf{2}$ & $1/2$&$0$  &$\omega^2$& $0$   \\
   $H_d^-$ & $\textbf{1}$ &$\textbf{2}$ & $1/2$&$-1$  &$\omega^2$& $0$   \\
   \hline
  $H_d^{'0}$ & $\textbf{1}$ &$\textbf{2}$ & $1/2$&$0$  &$\omega^2$&$\pi R/2$     \\
$H_u^{'0}$ & $\textbf{1}$ &$\textbf{2}$ & $-1/2$&$0$  &$\omega^2$&$\pi R/2$     \\
\end{tabular}
\caption{\label{ta:hh} Higgs fields located at the different branes. The $H^-_{u,d}$ are the physical Higgses and are renamed as $H_{u,d}$ (see discussion in the main text).}
\end{table}

These Higgses couple to the KK modes as
\begin{equation}
\begin{split}
\mathcal{L}_Y\sim &\ H_{u}^-\Big[(Q_{LKK}^-)^\dagger u_{RKK}^0+(Q_{LKK}^0)^\dagger u_{RKK}^+\Big]\\
&+ H_u^0\Big[(Q_{LKK}^0)^\dagger u_{RKK}^0+(Q_{LKK}^+)^\dagger u_{RKK}^+ +(Q_{LKK}^-)^\dagger u_{RKK}^-\Big]\\
& + H_u^+\Big[(Q_{LKK}^+)^\dagger u_{RKK}^0+(Q_{LKK}^0)^\dagger u_{RKK}^-\Big]\\
&+ H_u^{'0}\Big[(Q_{LKK}^0)^\dagger u_{RKK}^0+(d_{LKK}^+)^\dagger Q_{RKK}^+ +(d_{LKK}^-)^\dagger Q_{RKK}^-\Big]\\
&+ u\leftrightarrow d
\end{split}
\end{equation}

\begin{figure}
 \centering
\includegraphics[scale=0.3]{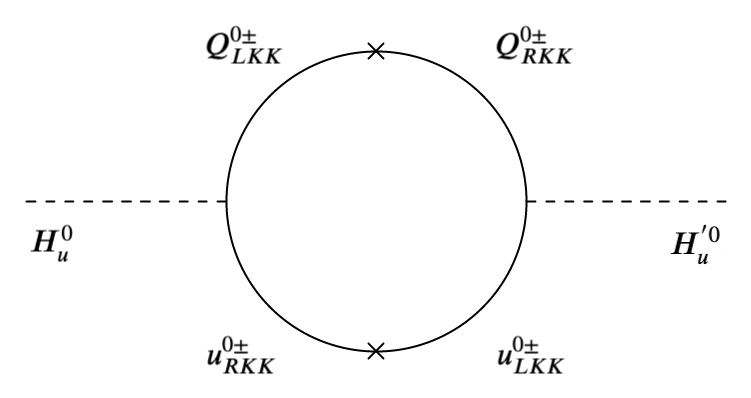}
		\caption{Loop diagram generating isoneutral Higgs doublet mass mixing. A similar diagram 
		for down type Higgs doublets is obtained by changing $u\rightarrow d$.}
		\label{fig:loop}
\end{figure}

Integrating out the KK modes would generate the mass terms from the diagram in fig. \ref{fig:loop}, which would generate the 
mass mixing term (with the compactification scale acting as a natural cutoff)
\begin{equation}
%\mathcal{L}_{HE}\sim \frac{1}{16\pi^2}M_{KK}^2\left(1+\ln \frac{M_{u,d}}{M_{KK}}\right)(H_{u,d}^{0})^\dagger H_{u,d}^{'0}
\frac{1}{16\pi^2}M_{KK}^2(H_{u,d}^{0})^\dagger H_{u,d}^{'0}
\end{equation}
where $M_{KK}^2\sim 1/R^2$ and we ignore the 5d Dirac masses for simplicity.
Although this mass will be a factor $\frac{1}{4\pi}$ smaller than $M_{KK}$,
it will be sufficiently large to deter either 
of the isoneutral Higgs doublets 
$H_{u,d}^{0}$ or $H_{u,d}^{'0}$ from developing a VEV.

On the other hand, the isocharged Higgses $H_{u,d}^{\pm}$ (the remaining parts of the isotriplets on the zero brane) 
do not receive such mass contributions from radiative effects, so they will readily develop VEVs. However, as only $\phi$ with a isocharge $+1$  gains a VEV, only the $H_{u,d}^-$ can generate effective Yukawa couplings. We rename these physically relevant Higgs doublets as 
$H_{u,d}\equiv H_{u,d}^-$.

\end{document}